# Beyond the Grid: Cost, Carbon, and Capital Requirements of On-Site Power Technologies for AI Data Centers

Eliseo Curcio

## Abstract

Interconnection queues, not electricity prices, now govern where data centers can be built, and the standard levelized-cost comparison answers a question no developer faces: it assumes a load profile, freezes the grid price while modeling the demand that moves it, and quotes busbar costs a facility cannot buy. This paper evaluates nine on-site supply technologies against a delivered grid whose price is endogenous to projected data-center demand, on a complete-site basis that retains standby charges, with measured GPU training load, delivered fuel prices, production-pathway carbon, and statutory 45V and 48E incentive mechanics. Nothing beats the wire: gas combined cycle produces at $47/MWh but costs about $114 per megawatt-hour of complete site energy against a $92 grid; four-hour storage is physically capped near 18 percent of annual energy and, charged at the margin, dirtier than the grid; hydrogen from grid-priced power fails on cost and carbon together. An investment inversion converts these findings into capital terms: conversion-hardware learning buys nothing, because free hardware still exceeds the grid for every low-carbon arm, while global electrolyser deployment on sited sub-$20/MWh power brings PEM hydrogen power to about 2.2 times the grid at $300 billion and 1.9 times at $1 trillion (2.7 and 2.3 for the hydrogen engine), with a carbon reduction of roughly 85 percent—6.8-fold—against grid-power production. Grid parity is not purchasable at any budget. On-site supply is an access and depth product; most current investment targets the wrong term.

## 1. Introduction

Estimates place United States data-center electricity consumption at 649 TWh in 2030 in the reference case of the national laboratory assessment that anchors this study [1], up from 192 TWh in 2024 [1], with a plausible range of 521 to 843 TWh [1]. The prior edition of the same series established the 2014-2023 historical record [2]. Interconnection is the binding constraint on serving that growth: connection queues in the largest markets run five to ten years [3], a majority of queued capacity may never energize [3,4], and of roughly sixteen gigawatts of project capacity scheduled to begin operating in 2026, only about five gigawatts was under construction at the start of the year [5,6]. The gap between announced and delivered capacity is now the central planning fact of the industry, and it has pushed developers toward generation behind their own meter.

The scale of the demand shock deserves quantification before technologies are compared, because it sets both the size of the problem and the price of the default answer. In the anchor assessment's reference case, data centers reach 11.8 percent of United States electricity by 2030 [1]; converting consumption to capacity at the fleet's roughly fifty-percent average utilization implies about 148 gigawatts of data-center power demand in 2030 and 230 by 2040 in the reference case, and over 450 in the high case. Individual projects have grown to match: the largest facility in this study's scale analysis is one hundred megawatts, while announced campuses now reach ten gigawatts [7], and the marginal project driving the late-2020s buildout is a multi-gigawatt campus rather than an average facility. Meanwhile the delivered price of the default answer—the wire—is not standing still. Utilities recovering lumpy transmission investment from concentrated new load, gas burned at the margin to serve it, and capacity procured against a tightening reserve margin all flow into the tariff the data center pays. A comparison that freezes the grid at its 2026 price while modeling the demand that moves it is answering a question no developer faces.

The question this paper answers is which on-site technologies can carry that load, at what cost, with what carbon consequence, and—critically—what quantity of investment would change any of those answers. The last question is usually missing. A finding that a technology costs five times the grid today is a description; a decision requires knowing whether a billion dollars, a hundred billion, or no amount of money closes the gap, and through which term of the cost equation the money must flow.

Three methodological choices distinguish this study from the standard levelized-cost comparison. First, the load is measured, not assumed. Artificial-intelligence training load is a fast, high-amplitude process: in power telemetry collected from a 72-billion-parameter training run, 53.2 percent of samples exceed ±200 W/s of ramp. Rather than excluding slow generators outright, the analysis prices the battery buffer needed to bridge each generator's response time, which turns out to cost single dollars per megawatt-hour and reframes ramp capability as a system cost instead of a categorical failure. Second, the grid benchmark is endogenous. Every prior comparison the author is aware of holds the grid price fixed while modeling a demand shock large enough to move it. Here, projected data-center consumption feeds power-sector gas burn through a supply elasticity and escalates transmission and network charges toward a cap, so the delivered grid price in the reference scenario rises from \$74/MWh in 2026 to \$92/MWh in 2030 as a consequence of the modeled demand itself. Third, comparisons are made on a complete-site basis. A generator's busbar cost is not the cost of powering a data center: a facility that generates on site while retaining the grid for peaks, outages, and backup continues to pay capacity and network charges. Only the energy component of the grid bill declines with on-site share in this model; the fixed components persist unless the firm grid relationship is demonstrably shed. This single accounting choice reverses the most publicized finding in the field, as Section 4 shows.

The paper also inverts the cost model rather than merely evaluating it. For each technology it solves for the parameter values—hardware capital cost, fuel price, electrolyser electricity price—at which the technology reaches the grid, tests those values against credible floors from the published literature, and then converts the reachable ones into cumulative deployment investment through Wright's law. The result is a decision table in dollars: what money buys, at what price per megawatt-hour of cost reduction, and where money stops working.

## 1.1 Relation to existing practice

Three bodies of work bear on the question and each stops short of it. Levelized-cost surveys [8,9] provide the parameter backbone of any comparison, including this one, but their headline figures are busbar costs at survey-standard capacity factors, agnostic to the load being served, the grid relationship retained, and the fuel logistics behind the plant gate; applied unadjusted to a data-center siting decision they produce the storage-below-grid and gas-below-grid readings that this paper's accounting reverses. Data-center energy assessments, of which the national-laboratory series is the most rigorous [1,2,10], project consumption bottom-up from equipment shipments but treat supply as exogenous, so they quantify the demand shock without pricing its consequences for the tariff that shock arrives on. And the hydrogen techno-economics literature, including the author's prior work on production cost and delivery infrastructure [11], optimizes the fuel chain in dollars per kilogram—a currency in which a data-center operator cannot transact. The contribution here is the join: measured load feasibility, an endogenous grid benchmark, complete-site accounting, delivered fuels, statutory policy mechanics, and an investment inversion, in a single model whose every table reproduces from released code, expressed throughout in dollars per megawatt-hour of delivered site energy, which is the only currency the decision is made in.

Two prior findings from the author's related studies feed directly into the model rather than sitting beside it. The green-hydrogen production analysis [11] established that electrolytic hydrogen requires renewable power in the twenty-to-thirty-dollar band to reach competitive production cost, and that finding reappears here from the demand side: the data-center inversion independently lands on the same power-price band as the condition for hydrogen electricity at twice grid, which is the kind of agreement between two unrelated derivations that deserves more weight than either alone. And the measured GPU power-control work [12] established that training load can be actively shaped at the facility scale, which matters here because a facility that curtails training load under a power controller is, in those same hours, the natural buyer of the curtailed renewable energy that makes its hydrogen clean and cheap—the siting condition and the load-flexibility capability are the same asset viewed from two sides.

Section 2 describes the model, the fitted scenario axis, and the policy implementation. Section 3 presents the central cost and carbon results. Section 4 covers deployment configuration, penetration limits, and scale. Section 5 presents the investment inversion. Section 6 discusses implications and limitations, and Section 7 concludes. The model, a version-controlled revision log spanning fourteen revisions and six rounds of external review, and a workbook reproducing every table are released with the paper.

# 2. Model and data

## 2.1 Technology set and measured load

Nine supply technologies are compared against a delivered-grid reference: combined-cycle gas, gas peaking turbines, four-hour lithium-ion storage, solid-oxide fuel cells on pipeline natural gas and on renewable natural gas, proton-exchange-membrane fuel cells on hydrogen, hydrogen reciprocating engines, renewable natural gas reciprocating engines, and hydrogenated-vegetable-oil gensets. Capital, operating, efficiency, lifetime, and lifecycle-carbon parameters are drawn from published cost surveys [8,9], vendor deployment data [13,14], national-laboratory costings [15], and lifecycle-assessment literature [16,17], each with a declared boundary; the full parameter table with sources ships in the accompanying workbook. Each technology carries a minimum practical unit size—twenty megawatts for combined cycle, five for peakers, half a megawatt for engines, a tenth for fuel cells—and configurations below it return no result rather than a fictitious cost.

Load dynamics come from accelerator power telemetry recorded during reinforcement-learning training runs at three model scales on A100-class hardware. Facility load is constructed from the traces through a stated chain—accelerator share of IT load, server share of facility electricity [19], and a power-usage-effectiveness range for direct liquid cooling [18]—using a block bootstrap whose block length exceeds the dominant rollout period, so the sub-second autocorrelation that motivates the exercise survives the sampling. The operative statistic is that a majority of samples in the largest run exceed ±200 W/s. A generator whose response time is minutes cannot follow that load directly, but a data center does not ask it to: uninterruptible-power systems already sit between the IT load and the supply. The model therefore prices a buffer battery sized to absorb thirty percent of rated load for the generator's response time—five minutes for the slow class, one minute for the medium class. The energy requirement is small (0.025 kWh per kW for the slow class), nearly all the cost is power electronics, and the resulting adder is $0.9/MWh at high duty, rising to about $3/MWh at a hundred-percent swing assumption. Ramp capability is thus a priced constraint, not an exclusion; an earlier revision of this model that excluded slow generators outright overstated the constraint by two orders of magnitude and its results were withdrawn.

## 2.2 Cost model

Levelized cost follows the standard annuity form at a 7.7 percent weighted cost of capital [8] over technology-specific lifetimes, with fixed and variable operations, fuel at its delivered price, and carbon where a compliance price applies. The central case is the direct discounted cost with no calibration multiplier. A single multiplier fitted to reconcile the annuity against a published levered-return benchmark—used in early revisions of this model—cannot represent the different financing, maintenance, replacement, and stack-life profiles of fuel cells, engines, and turbines with one number, and sensitivity analysis showed it flipped two hydrogen conclusions on its own. It is retained only as a high-cost sensitivity, and the workbook marks every claim that changes sign between the two treatments.

Two accounting boundaries required particular care. Compliance carbon prices apply to covered stack emissions only—about 342 gCO2/kWh for combined cycle at its heat rate, using the federal natural-gas emission factor [20]—not to lifecycle intensity, which would charge an allowance price against upstream methane and manufacturing emissions no allowance covers; lifecycle intensity is retained separately for abatement comparison. And the carbon cost embedded in the grid energy price uses the marginal gas unit's stack rate, since gas sets the price, rather than the average lifecycle intensity of the whole mix, which confuses both a boundary and an aggregation.

Hydrogen enters at its delivered price, not the plant gate. Compression, on-site storage, site infrastructure, transport (trailer for merchant supply, zero for co-located production), and three percent losses add $1.15/kg before transport for co-located supply and about $2.65/kg for trailered supply. Storage is treated throughout as a system component rather than a generator: a four-hour battery cycling once daily over 350 days has a maximum discharge capacity factor of 16 percent under the survey-standard cycling schedule [9], its charging cost uses an incremental tariff (the marginal energy price plus a small demand adder) and its charging emissions use the matching marginal source, and it appears in supply tables only with its charging pathway named. An earlier revision that reported storage as a 90-percent-capacity-factor primary source at $62/MWh embodied 5.6 impossible cycles per day; the corrected figure at its physical ceiling is $118/MWh from incremental grid charging.

### 2.3 Scenario axis: fitted, then coupled

Scenario variables are fitted from observed data rather than imported from published outlooks, which are used only for validation. Grid carbon intensity is built bottom-up from a fitted generation mix: logistic curves on observed solar and wind shares, exponential decay on coal at its observed 3.6 percent annual rate, nuclear and hydropower flat where the record shows no trend, and gas as the residual, with lifecycle emission factors per source scaled once to reproduce the observed 2024 national intensity of 384 gCO2/kWh [21], with per-source lifecycle factors from the assessment literature [16]. The resulting decline of 3.7 percent per year through 2040 is an output of the fitted mix, not an assumed rate, and its dominant uncertainty—the solar logistic ceiling—is swept from thirty to sixty percent terminal share, moving 2040 intensity between 262 and 192 gCO2/kWh. Natural gas follows a mean-reverting fit to observed Henry Hub annual averages [22] with the long-run mean computed from the series; this is acknowledged as the weakest fit in the model, since five annual observations cannot express the liquefied-natural-gas export build that drives published forecasts higher, and the reference path should be replaced by a fundamentals model before the projections are used for trading purposes. The carbon price is fitted to observed regional allowance auction clearings—$13.00 in late 2021 rising to $35.00 by mid-2026 [23], a 21 percent annual trend—and damped logistically toward the federal social cost of carbon [24], reaching $119 per short ton — $131 per metric tonne — by 2040; the ceiling is a scenario device, not a market forecast, and the text and workbook say so. Hydrogen is not a scenario input at all: its price is derived from electrolyser capital under Wright's law [27] plus the scenario's electricity price, which ties the hydrogen story to the electricity story and removes the objection that the hydrogen conclusion is an artifact of an assumed hydrogen price.

Validation against published forecasts is reported as a result rather than used as an input. The fitted grid-intensity decline of 3.7 percent per year runs faster than the agency's near-term 1.4 percent figure and close to its 3.2 percent alternative [25], for a visible reason: the fitted solar logistic carries share from nine percent toward its ceiling while coal decays out, and disagreement with the output requires disagreement with that fit. The gas reference path sits below both major published outlooks at 2040—the honest consequence of a fit that cannot see the export thesis—while the high case, constructed as one standard deviation above the observed mean, moves toward a leading consultancy's 2030 figure [26], which is the only kind of agreement worth reporting. And the electrolysis investment curve integrates to about $2.9 billion per gigawatt over the agency's benchmark range against its published $3.6 billion, the difference being that learning front-loads the cost decline.

The four scenarios are then two drivers each. S1 (reference) uses the fitted gas path, no carbon price, the central mix, and $45/MWh electrolyser power. S2 (grid constrained) uses the high gas case, the slow-solar mix, $60/MWh power, and the high demand path. S3 adds the fitted carbon price to S1. S4 (hydrogen scales) is S1 with electrolyser power at $20/MWh from curtailed renewables and the production tax credit active.

Table 1 summarizes the derived drivers. Two properties are worth noting. Hydrogen's price now moves for structural reasons: it falls with electrolyser learning, falls further where the production credit applies, and rises when the credit's ten-year window closes for the modeled cohort—all mechanisms, not dials. And the grid carbon-intensity cases are named after the solar ceiling that generates them rather than after a decline percentage, which keeps the argument attached to the observable that drives it: a reviewer who disputes the 2040 intensity must dispute a logistic fitted to observed generation shares, not a cited rate.

| Driver, 2030 / 2040 | S1 Reference | S2 Constrained | S3 Carbon-priced | S4 H2 scales |
|---|---|---|---|---|
| Henry Hub gas, $/MMBtu | 3.26 / 3.03 | 4.77 / 5.35 | 3.26 / 3.03 | 3.26 / 3.03 |
| Carbon price, $/t metric | 0 | 0 | 52 / 131 | 0 |
| Grid CI, $gCO_2e$/kWh | 266 / 177 | 306 / 240 | 265 / 150 | 266 / 177 |
| Electrolyser power, $/MWh | 45 | 60 | 45 | 20 (curtailed) |
| Delivered $H_2$, $/kg | 7.68 / 6.9 | 8.5 / 7.7 | 7.68 / 6.9 | 3.81 / 3.9 |
| Delivered grid, $/MWh | 92 / 98 | 150 / 161 | 113 / 120 | 92 / 98 |
| DC demand, TWh | 649 / 1,008 | 850 / 2,012 | 649 / 1,008 | 649 / 1,008 |

*Table 1. Derived scenario drivers. Gas, carbon, and grid intensity are fitted from observed series; hydrogen is computed from electrolyser learning plus the power price; the delivered grid price is endogenous to the demand path shown in the last row.*

## 2.4 Endogenous demand

The coupling that distinguishes this model runs from demand to price. Incremental data-center consumption above the 2026 level is served at the margin largely by gas; the model routes sixty percent of it into power-sector gas burn and moves the gas price along a constant supply elasticity of 0.35. Peak-demand growth escalates the capacity and network components of the delivered grid price at five times the proportional rate—a deliberately super-proportional judgment reflecting the lumpy, front-loaded transmission buildout that concentrated new load requires—capped at 1.6 times the 2026 charge. Both elasticities are judgment parameters, flagged as such, and the proportional case is one line to run. The consequence is that the reference-case delivered grid price rises from $73.8/MWh in 2026 to $91.8 in 2030 and $98.4 by 2040 as the modeled demand builds, and the constrained scenario reaches $150 by 2030 from its own demand path rather than from hand-picked adders. Every on-site technology in this paper competes against that moving benchmark, not against a frozen 2026 tariff.

## 2.5 Policy implementation

The Section 45V clean-hydrogen production credit is implemented as current guidance specifies [28]: tiered on the carbon intensity of hydrogen through the point of production (delivery emissions excluded), with the full rate requiring less than 0.45 kgCO2e/kg and the 1.5–2.5 kg band receiving one quarter of it; the maximum is the inflation-indexed 2025 base of $0.637/kg times the five-fold prevailing-wage multiplier, indexed forward; eligibility requires construction start before 2028; and the credit runs ten years from the placed-in-service date rather than ending for all facilities in a single year. Under these rules the S4 co-located pathway, at 1.65 kgCO2e/kg production intensity, earns $0.88/kg in 2030—not the full $3—and delivered S4 hydrogen is $3.81/kg. The Section 48E investment credit [29] applies only to facilities with an anticipated greenhouse-gas emissions rate of zero or below, which excludes natural-gas fuel cells; its phase-out begins at the later of 2032 or the year power-sector emissions reach one quarter of their 2022 level, a threshold the fitted mix does not reach before 2040, so the thirty-percent credit persists through the modeling horizon for storage and hydrogen fuel cells.

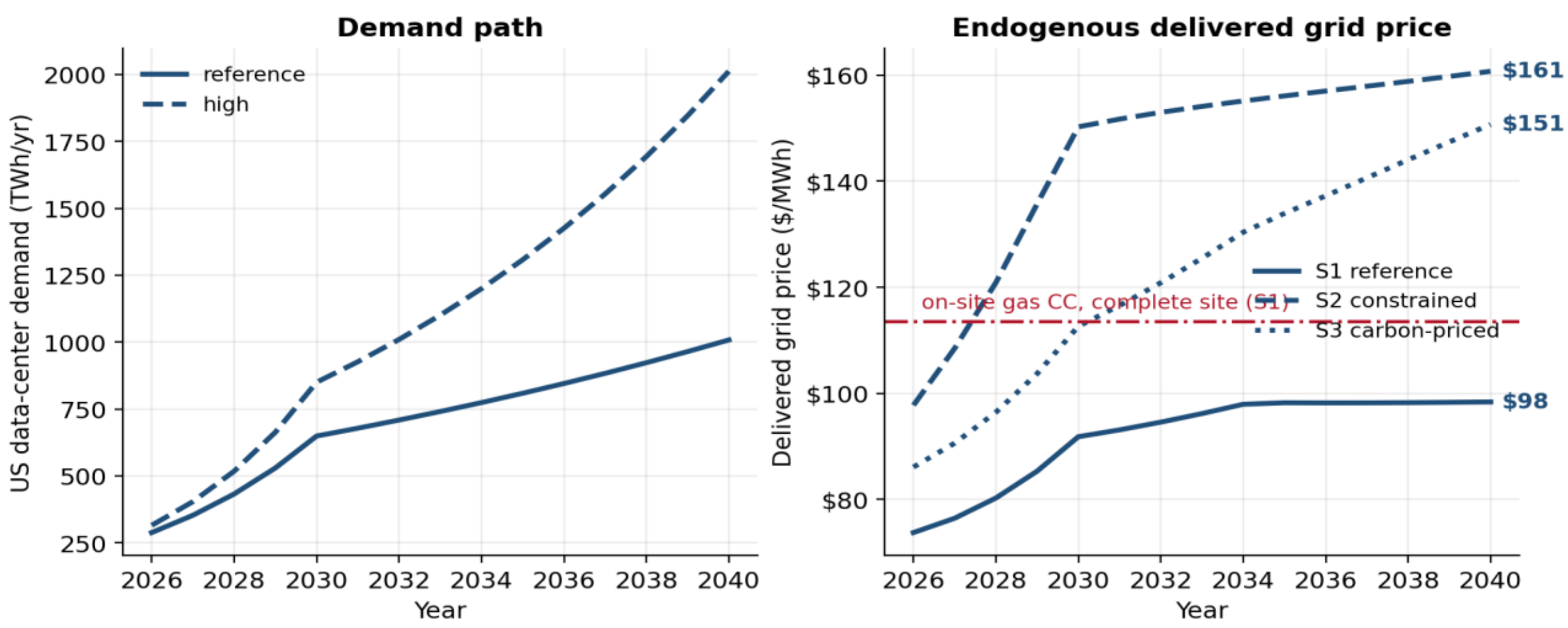

*Figure 1. The model's central novelty: projected data-center demand (left) and the delivered grid price it produces (right). The benchmark every technology competes against is moved by the demand it serves; the on-site gas complete-site line is shown for comparison.*

# 3. Central results: cost and carbon

## 3.1 The complete-site comparison

Table 2 reports 2030 system costs under the reference scenario at each technology's physically sustainable capacity factor, inclusive of the ramp buffer, against the demand-inflated delivered grid of $92/MWh. Two columns matter: the busbar cost, and the complete-site cost when the facility retains its firm grid relationship.

| Technology | CF | System LCOE ($/MWh) | Complete site @100% ($/MWh) | Lifecycle gCO2e/kWh |
|---|---|---|---|---|
| Delivered grid (endogenous) | 90% | 92 | 92 | 266 |
| Gas combined cycle | 95% | 47 | 114 | 431 |
| Gas peaker | 95% | 58 | 125 | 611 |
| SOFC, natural gas | 95% | 82 | 148 | 373 |
| BESS 4h (incremental charging) | 16% | 118 | 108 at its 17.8% cap | 557 |
| SOFC, renewable natural gas | 95% | 179 | 246 | 85 |
| RNG reciprocating engine | 95% | 194 | 260 | 100 |
| HVO genset | 95% | 279 | 346 | 189 |
| PEM fuel cell, delivered $H_2$ | 95% | 490 | 556 | 915 |
| $H_2$ reciprocating engine | 95% | 609 | 676 | 1,144 |

*Table 2. 2030 reference-scenario costs at sustainable capacity factors. The complete-site column blends on-site supply with retained grid capacity and network charges; the storage entry is priced at its 17.8% annual-energy cap and hydrogen from grid-priced power exceeds coal-fired carbon intensity.*

The headline is in the gap between the second and third numeric columns. Gas combined cycle produces electricity for $47/MWh, half the delivered grid—and this is real, driven by a fuel price near $3.26/MMBtu against a delivered tariff carrying $46/MWh of capacity and network charges. But a data center cannot buy the busbar number. A right-sized combined-cycle plant supplying one hundred percent of annual site energy, with the grid retained for peaks, outages, and the reserve margin the on-site unit cannot provide, costs about $114/MWh of total site energy—$21 to $22 above the grid, a gap that persists through 2040 because the fixed grid components that create it persist. Reaching parity requires demonstrably shedding the firm grid relationship, which is a contractual and reliability question, not a generation-cost question. Nothing in the technology set beats the delivered grid on a complete-site basis, in any scenario, in any year of the horizon.

The scenario axis reshapes the ranking without reordering it. Under the constrained scenario the grid reaches $150/MWh by 2030 and $161 by 2040, driven by its own demand path; the complete-site gas figure rises too, on the same gas price, but by less, and the premium narrows to roughly 11–13 percent—about $20/MWh in 2030 and $18 in 2040—the closest approach in the scenario space, and it occurs precisely where the wire may not be available at any price, which is the point. Under the carbon-priced scenario the delivered grid carries the marginal gas unit's allowance cost and reaches $113/MWh in 2030; on-site gas pays the same allowance on its own stack and gains nothing relative, while the clean arms gain $20 of headroom that still leaves the nearest of them eighty dollars short. Carbon pricing at fitted-trajectory levels reranks nothing in this set; it is not the instrument that makes any of these technologies competitive, and analyses that make it one are assuming prices the observed auction series does not support. The hydrogen scenario changes hydrogen only, and Section 5 quantifies exactly what that change costs to reproduce elsewhere.

Storage illustrates why physical accounting matters more than headline levelized cost. At its cycling ceiling the four-hour battery delivers energy at $118/MWh from incremental grid charging—competitive-looking against the low-carbon arms—but it stores rather than generates, tops out near eighteen percent of annual site energy, and its charging electricity, priced at the margin, is made by the marginal gas unit, giving 557 gCO2e/kWh delivered: more than twice the average grid it displaces. Storage is a reliability and arbitrage service. It is not a decarbonization instrument when charged from a gas-margin grid, and it is not a primary energy source at any price.

The hydrogen rows carry the study's sharpest result. Hydrogen made by electrolysis on scenario-priced electricity costs $7.68/kg delivered and produces power at $490/MWh through a fuel cell—5.3 times the grid—while emitting 915 gCO2e/kWh, because 55 kWh of 266-gram electricity is embodied in every kilogram. Grid-electrolytic hydrogen fails on cost and carbon simultaneously; the expensive option and the dirty option are the same option. Under the S4 pathway—co-located electrolysis on curtailed renewable power at $20/MWh with the correctly tiered credit—delivered hydrogen falls to $3.81/kg, fuel-cell power to $257/MWh, and intensity to roughly 135 gCO2e/kWh. (Section 5 quotes a slightly higher co-located starting figure of $347/MWh because its learning curve deliberately starts from the conservative ex-China electrolyser cost of $2,300/kW, whereas the scenario's derived hydrogen price embeds a $1,700/kW electrolyser; the $90/MWh difference is the price of that conservatism, stated rather than hidden.) The pathway sharply improves both dimensions at once—though only the carbon comparison crosses the grid benchmark—which is why Section 5 treats siting as a precondition of the investment case rather than a refinement of it.

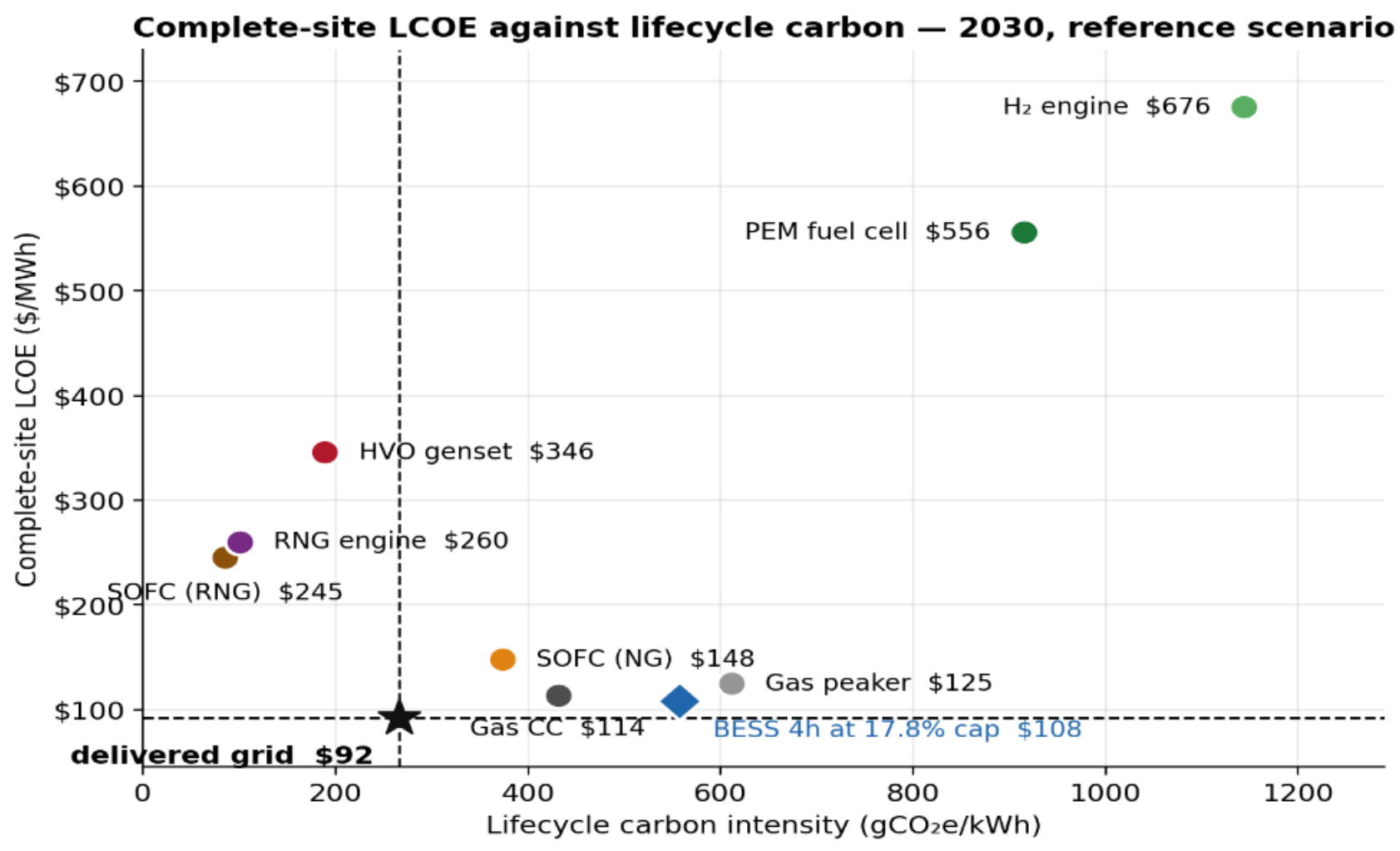

*Figure 2. Complete-site LCOE against lifecycle carbon intensity, 2030 reference scenario. Each point is a technology supplying the site at its sustainable maximum, blended with retained grid charges; storage (diamond) is plotted at its 17.8 percent annual-energy cap. No technology occupies the cheaper-and-cleaner quadrant relative to the delivered grid.*

## 3.2 Where the cost sits

Decomposing levelized cost explains which investments can work before any are priced. For every low-carbon fuel arm, fuel dominates: 94 percent of PEM cost, 95 percent of the hydrogen engine, 92 percent of HVO, 84 percent of the RNG engine, 66 percent of the RNG solid-oxide cell. Hardware capital is two to three percent for the hydrogen arms and eight for the RNG engine; only the solid-oxide cells, at roughly a quarter, have a capital story at all. The free-hardware floor—fuel and operating costs with conversion-hardware capital set to zero—is the decisive statistic: $138/MWh for the RNG solid-oxide cell, $179 for the RNG engine, $271 for HVO, $479 and $597 for the hydrogen arms—and every low-carbon arm's floor sits above the $92 grid. Free hardware moves single digits. The learning-curve case for funding fuel-cell and genset cost reduction, which dominates deployment-support discourse, is aimed at a term that does not bind.

## 3.3 Carbon accounting and abatement

With production-pathway carbon and marginal charging correctly assigned, the abatement table thins dramatically. Under the reference scenario, gas, the natural-gas fuel cell, storage, and both grid-hydrogen arms are dirtier than the grid they would displace and abate nothing. What remains: the RNG solid-oxide cell abates at roughly $482 per tonne, the RNG engine at $615, and HVO at $2,424—two and a half to thirteen times the $190 federal social cost of carbon. Under the curtailed-power scenario, clean hydrogen abates at $1,260 to $2,328 per tonne, against the reference grid whose carbon its co-located pathway does not displace hour-for-hour. No route in the set is competitive with generic power-sector abatement on price; the case for the survivors is depth and siting, not efficiency of public spending per tonne.

| Technology | Covered stack gCO2/kWh | Lifecycle gCO2e/kWh (S1) | Lifecycle (S4) | Abatement $/t (S1) |
|---|---|---|---|---|
| Gas combined cycle | 342 | 431 | 431 | dirtier than grid |
| Gas peaker | 584 | 611 | 611 | dirtier than grid |
| SOFC, natural gas | 312 | 373 | 373 | dirtier than grid |
| BESS 4h (incremental) | 0 | 557 | 557 | dirtier than grid |
| SOFC, RNG | 0 (biogenic) | 85 | 85 | 482 |
| RNG reciprocating | 0 (biogenic) | 100 | 100 | 615 |
| HVO genset | 0 (biogenic) | 189 | 189 | 2,424 |
| PEM fuel cell | 0 | 915 | 135 | dirtier / 1,260 (S4) |
| $H_2$ reciprocating | 0 | 1,144 | 169 | dirtier / 2,328 (S4) |

*Table 3. The two carbon boundaries. Covered stack emissions bear the compliance price; lifecycle intensity determines abatement. Conflating them—charging allowance prices against lifecycle intensity, or crediting storage and hydrogen with zero-carbon delivery—was among the errors corrected in review.*

# 4. Deployment configuration, penetration, and scale

## 4.1 Two ways to deploy, an order of magnitude apart

How a generator is sized matters more than which generator is bought. The model compares two configurations. In the first, a unit sized to the site's peak runs intermittently: supplying a quarter of a ninety-percent-utilization site's energy means operating at 22.5 percent capacity factor, and annualized fixed costs per megawatt-hour scale inversely with duty, so blended costs balloon at low shares. In the second, a unit sized to a fraction of peak runs continuously at its maximum capacity factor, priced at its own megawatt rating, and never pays the duty penalty. The difference is decisive: within a 25-percent tolerance of grid cost, the low-carbon fuel arms carry between four and forty-one percent of annual site energy when right-sized and continuous, versus at most a few percent when oversized and intermittent; the shares vary with site size because minimum practical unit sizes bind at the small end.

| Technology | Full-size, intermittent | Right-sized, continuous | Binding constraint on further share |
|---|---|---|---|
| Gas combined cycle | 100% (large sites only) | 100% (min. 20 MW unit) | standby charges keep site cost above grid |
| Gas peaker | 14–33% | 64–70% | fuel cost at high duty; carbon if priced |
| SOFC, natural gas | 0% | 38–41% | standby charges; carbon if priced |
| SOFC, RNG | 0% | 14–15% | regional RNG resource |
| RNG reciprocating | 0.1–1.1% | 13–14% | regional RNG resource |
| HVO genset | ≈4% | ≈9% | cost premium; commodity fuel |
| PEM fuel cell | <1% | ≈5% | delivered hydrogen price |
| $H_2$ reciprocating | <1% | ≈4% | delivered hydrogen price |
| BESS 4h | ≈17.5% | ≈17.5% | physics: 4 h × 1 cycle/day |

*Table 4. Largest share of annual site energy servable within 25 percent of grid cost, by deployment configuration, across 1–100 MW sites. Right-sizing—not technology choice, not facility scale—is the dominant free lever.*

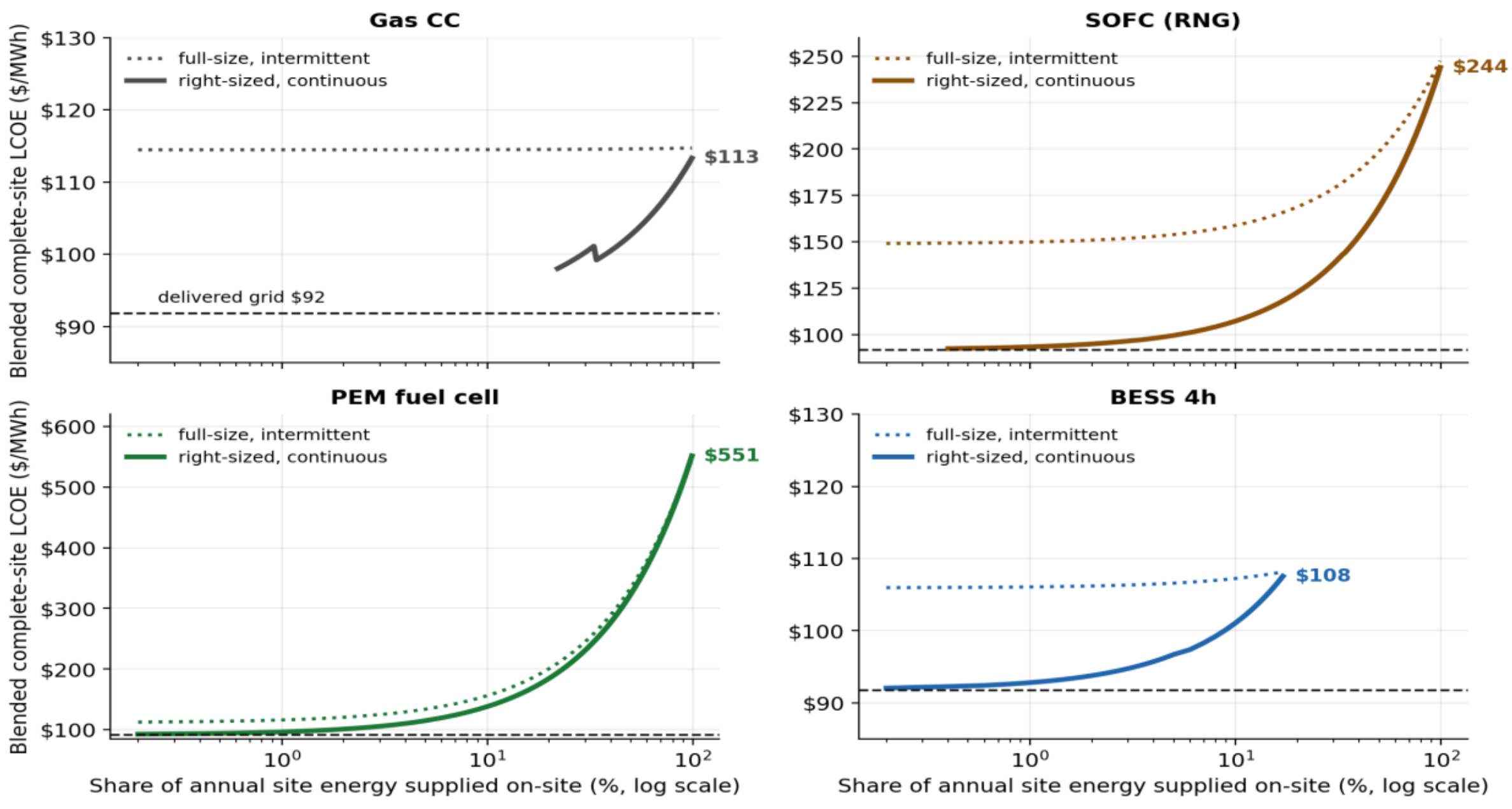

*Figure 3. Blended complete-site LCOE against on-site share, one panel per technology so the configuration gap is legible; note the differing vertical scales. Solid: right-sized continuous. Dotted: full-size intermittent. Dashed line: the $92 delivered grid. Gas CC's right-sized curve begins at the share its 20 MW minimum unit permits.*

The mechanism behind the gap deserves stating because the intuition runs the other way. One expects a technology at five times grid cost at full duty to sit near grid cost at a small share, since the grid carries the rest. It does not, in the oversized configuration, because annualized capital and fixed operations per megawatt-hour scale inversely with capacity factor: as the share falls, the unit's own cost rises fast enough to cancel the dilution, and the blended curve is nearly flat at a high level instead of falling toward the grid. The right-sized configuration escapes because the unit never leaves its maximum duty; its blended cost is a straight line between the grid price and the unit's full-duty cost, and the affordable share is simply where that line crosses the tolerance. The practical rule for a developer follows directly: never buy capacity you will not run, and if partial supply is the goal, shrink the machine rather than the schedule.

Two caveats attach to the right-sized result. A fractionally sized unit run flat out has no reserve headroom, so it is an energy product and cannot double as the site's backup; a facility wanting both pays for both. And minimum practical unit sizes bite at the small end: a combined-cycle plant cannot be right-sized to a ten-percent share of a hundred-megawatt site because twenty megawatts is the smallest unit that exists, which is why its right-sized column begins at the twenty-five-percent share.

## 4.2 Scale and the pilot-to-fleet gap

Sweeping facility size from one to one hundred megawatts moves costs remarkably little for the modular technologies—zero to four percent—because fuel cells, engines, and battery containers are factory-built and their unit economics are nearly scale-flat. Only rotating thermal plant carries a large small-scale penalty. What scale does move is the fuel chain. A single hundred-megawatt site running entirely on renewable natural gas consumes 6.4 trillion Btu per year against a New York resource assessed at 47 to 147 trillion Btu per year [30]; the state's entire resource supports seven to twenty-three such facilities, or several hundred at one megawatt. Hydrogen has no resource ceiling, but one hundred-megawatt site on hydrogen engines requires 59,000 tonnes per year, and 169 such facilities would consume the entire United States merchant hydrogen market, which is committed to refining and ammonia. The pilot-to-fleet gap for these technologies is therefore a fuel-supply gap, not an engineering gap: every arm demonstrates successfully at one to ten megawatts, and what fails at fleet scale is the supply chain behind it. This inverts the usual reading of technology-readiness evidence, in which a successful pilot is treated as the hard part.

The scale ladder itself deserves one honest note. The industry's marginal project is no longer a hundred megawatts: announced campuses now reach ten gigawatts, two orders of magnitude beyond this study's largest modeled facility. At that scale, on-site gas ceases to be gensets and becomes a dedicated plant with its own interconnection, and the resource arithmetic above binds proportionally harder. Extending the ladder to gigawatt-class facilities is the most important piece of future work this model points to.

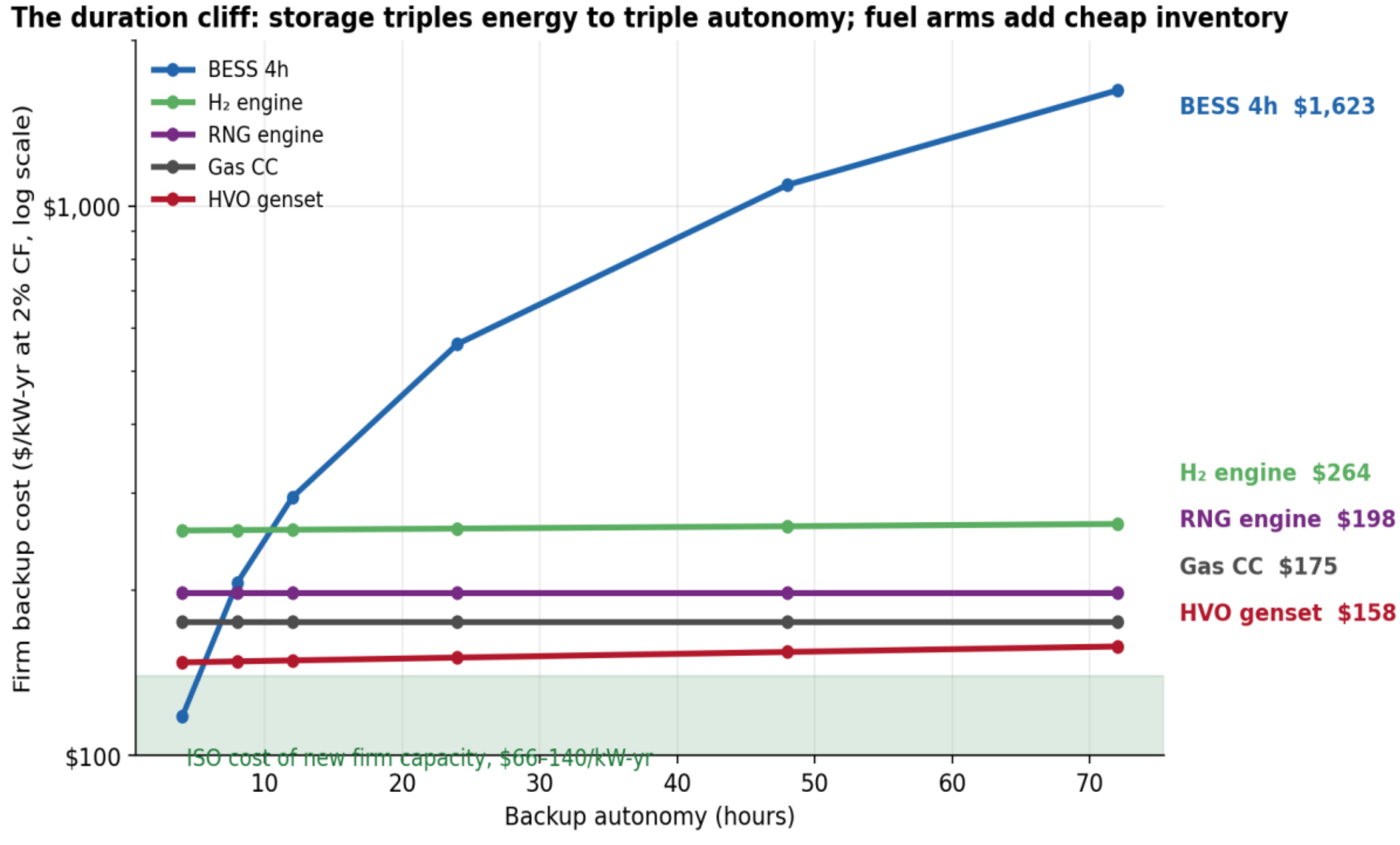

*Figure 4. Firm backup cost against autonomy, log scale. Storage scales its energy capacity with duration and climbs an order of magnitude; the fuel arms add vessel inventory at a nearly flat marginal cost. The shaded band is the system's own price of firm capacity, $66–140/kW-yr.*

## 4.3 Backup duty

Backup duty inverts the economics of the primary-supply comparison because the binding quantity changes from energy to stored autonomy. Extending a battery from four to twelve hours of ride-through means tripling its energy capacity, which is most of its cost; extending a hydrogen engine's autonomy means adding vessel inventory at thirteen-odd dollars per kilogram of storage, a far cheaper margin; and a pipeline-fed gas plant adds nothing but is only as firm as its pipeline, which winter events have shown is not fully firm. The model therefore prices each technology's autonomy scaling explicitly, netting out the working storage already embedded in the delivered hydrogen price so backup vessels are not double-counted—a correction from review, since the four-hour working inventory was previously charged twice.

In pure backup service at two percent capacity factor and twelve hours of autonomy, selected annualized costs include \$175/kW-year for combined cycle, \$258 for hydrogen engines, and \$295 for storage, against an independent-system-operator cost of new firm capacity of \$66 to \$140/kW-year. On-site backup costs one and a half to four times the system's own price of firmness. It is bought for availability during the interconnection queue and for outage insurance, not for economics—and that residual, the option value of being able to energize at all against queues in which most capacity never connects, is the quantity a developer must actually underwrite. The literature leaves it implicit; it should be priced explicitly, because after capacity-market credit it is the only remaining justification for most of the technology set.

### 4.4 Robustness: what six review rounds moved, and what they did not

The model underwent six rounds of external technical review across fourteen revisions, and reporting what changed is itself evidence about which conclusions deserve weight. Five findings survived every revision: the delivered grid wins on a complete-site basis; four-hour storage is a capped service, not a source; grid-electrolytic hydrogen fails on cost and carbon together; right-sizing dominates oversized intermittent operation; and clean co-located hydrogen lands at two to three times the grid, not at parity. Three findings from early drafts did not survive and were withdrawn: that storage supplies ninety-percent-capacity-factor power below the grid price (an impossible 5.6 cycles per day, traced to an anchoring error—corrected, storage delivers at \$118/MWh at its 16 percent cycling ceiling, and the \$108 figure appearing in Table 2 is the blended cost of a site taking storage at its 17.8 percent annual-energy cap with the grid supplying the rest); that half the technology set is infeasible on ramp dynamics (the buffer costs one dollar per megawatt-hour, not infinity); and that partial penetration cannot be economic (true only for the oversized configuration the early model happened to test). One finding flips with a single methodological choice and is reported as conditional throughout: whether hydrogen could ever have reached parity on hardware cost alone depended on the calibration treatment in earlier revisions; under the final direct-cost central case the free-hardware floors of all low-carbon arms exceed the grid unconditionally, and the workbook prints the superseded calibrated case for the record. The distinction between the five, the three, and the one is the paper's honest confidence interval.

Parameter sensitivity concentrates in known places. The solar-ceiling sweep moves 2040 grid intensity by seventy grams and the abatement denominators with it, without reordering the abatement ranking. The network-escalation judgment moves the 2030 reference grid between \$74 (proportional) and \$92 (super-proportional, central); the complete-site gas gap to the grid persists in both, at different absolute levels. The buffer swing assumption moves the ramp adder between one and three dollars—noise at the scale of the gaps involved. And the single most consequential input in the entire model is not a technology parameter at all but the electrolyser electricity price, which controls hydrogen's cost and carbon simultaneously and is set by siting, not by any market the model forecasts.

## 5. The investment inversion: what money buys, in $/MWh

The preceding sections establish present costs. This one asks the decision question: how many dollars of global deployment investment reduce each technology's levelized cost to a target, and where money stops working. Deployment investment operates through Wright's law—cost falls a fixed fraction per doubling of cumulative installed capacity—so a capital-cost target converts to gigawatts to deploy and thence to cumulative dollars, integrating the declining cost curve rather than multiplying endpoint price by endpoint volume. Learning rates come from the model's parameter set; installed bases come from agency assessments for electrolysis [13,14] (about five gigawatts installed globally at the start of 2026 [13], against 58 gigawatts per year of manufacturing capacity [14], so manufacturing is not the constraint—deployment is) and from flagged estimates for the conversion technologies. Every value in this section reproduces from the investment_inversion module released with the model; the electrolyser learning curve starts from the ex-China installed cost of $2,300/kW, which is deliberately conservative given the roughly $900/kW Chinese cost point already below most of the curve. The reference-scenario grid at $92/MWh and the complete-site gas alternative at $114 are the targets.

### 5.1 The hardware channel is closed

The first table in the analysis runs every budget from ten billion to three trillion dollars into each technology's own conversion hardware and reports the resulting system cost alongside the free-hardware floor—the best any hardware budget can ever achieve. The floors settle the question: $481/MWh for the PEM fuel cell, $597 for the hydrogen engine, $179 for the RNG engine, $271 for HVO, $140 for the RNG solid-oxide cell. Every low-carbon arm's floor exceeds the grid, because fuel does. Their budget columns barely move from ten billion to three trillion; the money saturates against a term it cannot touch. The quantified recommendation is blunt: the correct deployment-support budget for fuel-cell and genset hardware cost reduction, measured by its effect on the affordability of the electricity produced, is zero.

| Technology | LCOE now | $100 bn | $1,000 bn | $3,000 bn | Free-hardware floor |
|---|---|---|---|---|---|
| Gas combined cycle | 47 | 47 | 47 | 47 | 29 |
| Gas peaker | 59 | 58 | 58 | 57 | 44 |
| SOFC, natural gas | 84 | 64 | 56 | 53 | 42 |
| SOFC, RNG | 181 | 162 | 153 | 150 | 138 |
| RNG reciprocating | 194 | 191 | 189 | 189 | 179 |

| Technology | LCOE now | $100 bn | $1,000 bn | $3,000 bn | Free-hardware floor |
|---|---|---|---|---|---|
| HVO genset | 279 | 278 | 278 | 277 | 271 |
| PEM fuel cell | 491 | 484 | 483 | 482 | 479 |
| $H_2$ reciprocating | 610 | 606 | 604 | 603 | 597 |

*Table 5. System LCOE ($/MWh) after cumulative global deployment investment in each technology's own conversion hardware, with the free-hardware floor. Every low-carbon arm's floor exceeds the $92 grid: the hardware channel is closed at any budget.*

## 5.2 The fuel channel works, to a floor

For the hydrogen arms the fuel price is itself capital—electrolyser plus electricity—so electrolyser deployment is the hydrogen learning investment, and it is the one channel in the study where money moves levelized cost at scale. Table 6 gives the result in the study's currency.

| Electrolyser power price | LCOE now | $100 bn | $300 bn | $1,000 bn | $3,000 bn |
|---|---|---|---|---|---|
| PEM fuel cell @ $45/MWh (grid-ish) | 486 | 372 | 339 | 311 | 294 |
| PEM fuel cell @ $30/MWh | 435 | 321 | 288 | 260 | 243 |
| PEM fuel cell @ $20/MWh curtailed + 45V | 347 | 233 | 199 | 172 | 155 |
| $H_2$ engine @ $20/MWh curtailed + 45V | 429 | 287 | 245 | 211 | 189 |

*Table 6. PEM and hydrogen-engine system LCOE ($/MWh) after cumulative global electrolyser deployment investment, by electricity price. The grid is $92/MWh; the on-site gas alternative $114. The curtailed rows use the co-located boundary—on-site transport, tiered 45V—which is the boundary the pathway implies; the $45 rows are merchant grid-power production, which is why Table 2's S1 value of $490 (trailered merchant supply) sits above this table's $45/MWh row. The co-located starting value of $347 exceeds the S4 scenario figure of $257 because this section's learning curve starts from the conservative $2,300/kW ex-China electrolyser cost rather than the scenario's $1,700/kW.*

Three readings. First, money without siting fails: three trillion dollars of electrolyser deployment at $45/MWh power still leaves fuel-cell electricity at $294/MWh—3.2 times the grid—and at 915 $gCO_2e/kWh$, because no deployment budget reduces the electricity term. The purchase would be an expensive way to burn gas twice. Second, money with siting works to a floor: one trillion dollars at $20/MWh curtailed power on the co-located boundary brings the fuel cell to $172/MWh—about 1.9 times the grid and 1.5 times the complete-site gas alternative—at roughly 135 $gCO_2e/kWh$. Third, grid parity is not on the curve at any budget: the delivered fuel budget at parity is $1.05/kg, while the cheapest hydrogen anywhere on the deployment curve—$300/kW electrolyser capital, $20 power—is $2.82/kg delivered before credit, and the credit closes at most $0.88 of the gap for its ten-year window. The answer to how much money buys hydrogen parity is: no amount. The purchasable frontier is the two-to-three-times band, at roughly $300 to $1,000 billion of cumulative global deployment—of which any single buyer or national program funds a sliver, since learning is a global commons—conditional on sited sub-$20 power. A useful cross-check: the international agency prices 420 gigawatts of electrolysis at more than $1,500 billion [14], or $3.6 billion per gigawatt, against this model's curve integral of about $2.9 billion per gigawatt over the same range—the same order, lower here because learning front-loads the

decline. It is also worth stating that the Chinese installed cost of roughly $900/kW already sits below the point the ex-China curve reaches after half a trillion dollars: the geographic cost gap is presently worth more than the entire learning ramp.

### 5.3 What the money cannot buy, and the ranked answer

The two interventions must be priced separately, because conflating them overstates what capital buys. The combined pathway shift—moving from merchant grid-power production to co-located curtailed power at $20/MWh, with on-site logistics and the tiered credit that pathway earns—is worth $139/MWh to the fuel cell and $174 to the engine before a dollar of learning investment is spent; it is a siting and contracting outcome, not a purchase. Learning investment alone, measured at fixed $20 power, buys $114/MWh for the fuel cell and $142 for the engine over the FIRST hundred billion dollars—an order of magnitude more than any other channel in the study—but the return decays steeply along the curve, as Wright's law requires: the next two hundred billion buy a further $34/MWh, and the seven hundred billion after that only $27 more, landing the fuel cell at 1.9 and the engine at 2.3 times the grid at the trillion-dollar mark, from 2.2 and 2.7 at $300 billion. The first tranche is where the leverage lives. The natural-gas solid-oxide cell is the only conventional arm deployment meaningfully moves ($84 to $56/MWh at a trillion dollars), and its residual is the standby relationship, not cost. The renewable-natural-gas arms are cheap enough to matter today but capped near fifteen percent of site energy by a resource no budget expands. HVO has no learning channel at all, because its cost is a commodity fuel. And what no channel buys is the electricity price and the carbon that rides on it: siting on curtailment is a contracting decision that flips hydrogen's cost and carbon simultaneously and costs effort rather than capital. The ranked conclusion for a capital allocator: fund electrolyser deployment sited on curtailed power if depth beyond the RNG ceiling is the goal; fund capacity-market access for behind-the-meter assets, which is free and offsets $66 to $140/kW-year of the standby burden; price the interconnection-access residual explicitly; and put nothing into conversion-hardware cost reduction.

| Technology | LCOE now | Best @ $1,000 bn (sited) | Reduction per $100 bn | Gap to grid after | What closes the residual |
|---|---|---|---|---|---|
| Gas combined cycle | 47 | 47 | 0.0 | −45 (busbar) | shed the standby relationship — contract, not capital |
| SOFC, natural gas | 84 | 56 | 19.6 | −36 (busbar) | standby relationship; carbon exposure if priced |
| SOFC, RNG | 181 | 153 | 19.6 | +61 | RNG resource ceiling, ~15% of site energy |
| RNG reciprocating | 194 | 189 | 2.9 | +98 | RNG resource ceiling |
| PEM fuel cell (sited, $20 power) | 347 | 172 | 114 | +80 | delivery floor; pathway shift worth $139/MWh |
| $H_2$ engine (sited, $20 power) | 429 | 211 | 142 | +119 | delivery floor; pathway shift worth $174/MWh |
| HVO genset | 279 | 278 | 1.0 | +186 | none — fuel is a commodity, no learning channel |

*Table 7. The investment decision table. Hydrogen slopes are first-tranche values ($0→$100bn) at fixed $20/MWh power, pricing learning investment alone; returns decay along the curve (the $100–300bn tranche yields roughly $17/MWh per $100bn). The pathway-shift gain is stated separately. The residual column names what money cannot buy.*

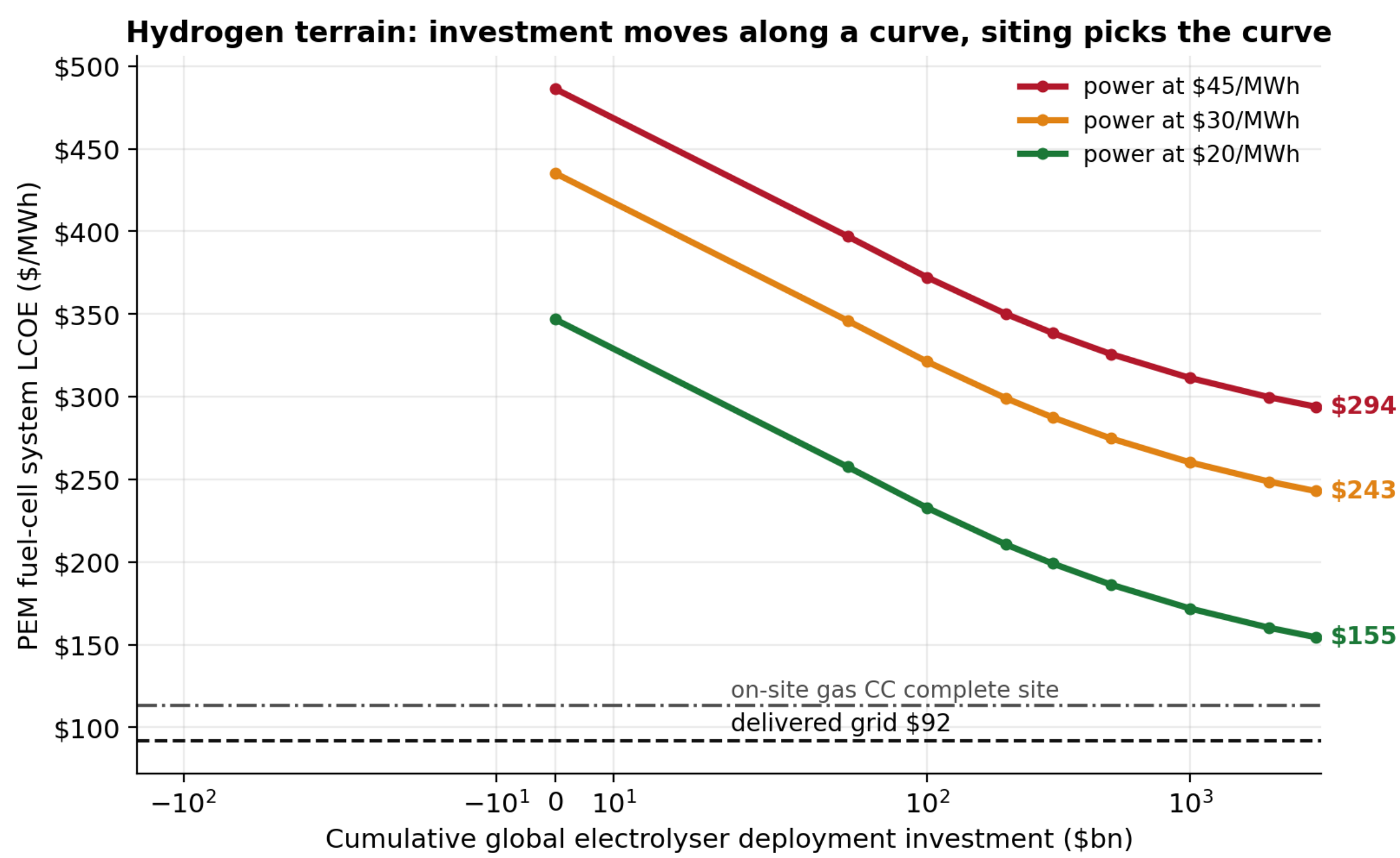

*Figure 5. The hydrogen terrain. Deployment investment moves cost along a curve; the power-price siting decision selects which curve. Grid parity lies below every curve at every budget.*

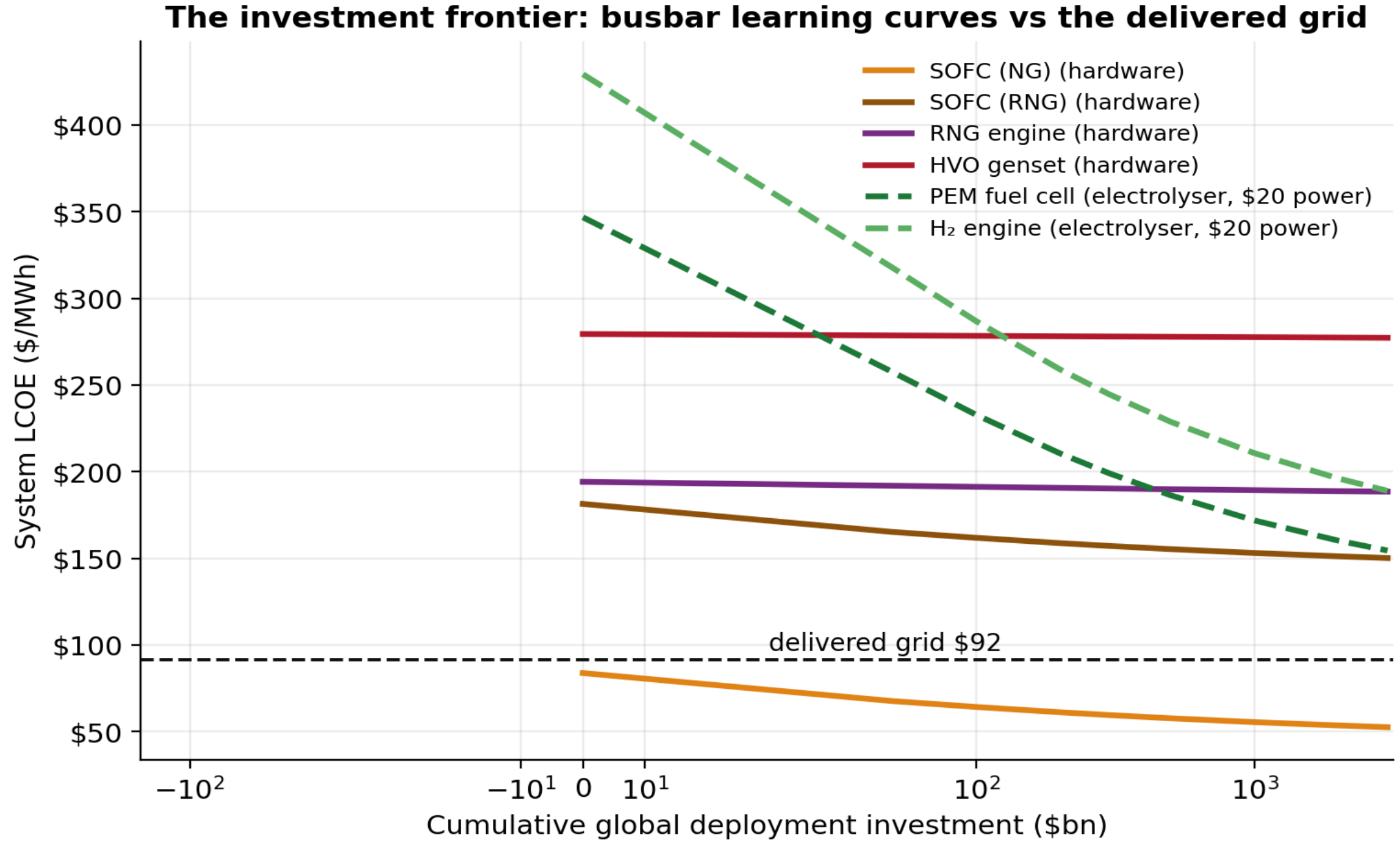

*Figure 6. The investment frontier across the technology set, in busbar terms. Hardware-learning curves for the fuel arms are nearly flat because fuel dominates their cost; the hydrogen electrolyser curves fall steeply from a high*

*start. The SOFC-NG busbar curve sits below the delivered-grid line, but its complete-site cost—with standby charges retained—remains above it (Table 2); no complete-site cost crosses the grid at any budget.*

# 6. Discussion

## 6.1 What the study changes

Three widely repeated claims do not survive the accounting in this model. The claim that on-site gas beats the grid confuses busbar cost with site cost; the $47 number is real and the $114 number is what a facility pays, and the difference is the grid relationship the facility keeps. The claim that batteries are the cheap clean option confuses a storage service with a generation source and average grid carbon with the marginal electricity a battery actually charges on. And the claim that hydrogen needs only cheaper electrolyzers or cheaper fuel cells confuses the two terms of its cost: the hardware term is two to three percent of the problem for the hydrogen arms, the fuel term is the problem, and the fuel term divides into a purchasable capital component and a non-purchasable siting component that also controls the carbon.

The endogenous-grid result also reframes the constrained scenario. When the wire itself reaches $150/MWh because of the demand it serves, on-site thermal generation stops being a premium product and becomes the marginal source of firm power for facilities the queue cannot reach—which is what is already observed in practice, where some jurisdictions now require large loads to bring their own generation rather than draw from the public grid. The model's contribution is to show that this outcome arrives through price alone, without any queue-failure assumption: the demand path of the constrained case is sufficient to price the grid to the point where self-supply is rational for facilities that can shed the standby relationship. Queue risk then operates on top of, not instead of, the price channel.

Conversely, two unfashionable conclusions strengthen. Right-sizing—running a smaller unit continuously instead of a big unit occasionally—is worth more than any technology substitution in the set and costs nothing. And the renewable-natural-gas arms, unglamorous and resource-capped, are the only routes that abate carbon today at three-digit rather than four-digit dollars per tonne; hydrogen's claim over them is depth beyond their ceiling, not price.

## 6.2 Implications by actor

For a data-center operator, the decision tree the model supports is short. If the local grid can serve the load, buy the wire: nothing beats it. If the queue cannot deliver on schedule, on-site gas is the least-cost bridge at roughly $114/MWh complete-site—accepting its 431 gCO2e/kWh—and the buffer architecture the measured load requires is a rounding error, not a barrier. If carbon commitments bind, renewable natural gas serves up to about fifteen percent of annual energy at three-digit abatement cost, and beyond that ceiling the only clean depth available is co-located hydrogen at approximately 2.8 times grid under the current S4 pathway, contingent on securing curtailed power—which the operator's own load flexibility makes easier to contract, since a facility that curtails training under a power controller is a natural counterparty for the same curtailed

energy. Storage belongs in every design for transients and outages and in no design as an energy source.

For a policymaker, the model prices the instruments. Capacity-market access for behind-the-meter assets is free and offsets a meaningful share of the standby burden. Production-side hydrogen support works only where the electricity is clean, and the statute's tiering already enforces this correctly—the analysis finds the tiered credit at $0.88/kg for the realistic co-located pathway, not the headline $3, and building policy expectations on the headline number overstates the support by a factor of three and a half. Deployment subsidies for fuel cells, engines, and gensets fail a quantitative test they are rarely subjected to: the free-hardware floor exceeds the grid, so the marginal abatement and affordability return on that spending is zero at any scale. And carbon pricing at levels the observed allowance market supports reranks nothing in this technology set; it is honest to say so rather than to model prices no auction has cleared.

For an investor, the asymmetry is the finding. The hydrogen channel offers the largest cost-reduction-per-dollar in the study and a floor well above the incumbent, which is the profile of an infrastructure platform play—value accrues to whoever owns the cheap-power siting and the offtake, not to whoever funds the learning, since learning is a global commons any deployment feeds. The renewable-natural-gas channel is the inverse: no learning to fund, immediate three-digit abatement, and a hard resource ceiling that makes it a portfolio of site-specific projects rather than a platform. The access residual—the option value of energizing at all against a queue where most capacity never connects—is the quantity the market is actually trading when it pays $114 for $92 power, and it remains unpriced as an explicit instrument.

## 6.3 Limitations

The model is a transparent scenario engine, not a dispatch model or an econometric market forecast, and its conclusions should be read at that resolution. The gas path is a five-observation mean reversion that cannot see the export build; the demand-to-price elasticities are judgment values, marked as such, with the proportional network case one line away; the allowance-price ceiling is a scenario device, since a compliance price and the social cost of carbon are different economic quantities; the generation-mix fit rests on four observation years with gas as the residual and no nuclear retirement or addition schedule; and the load traces come from A100-class hardware under one training regime, with the ramp statistic hard-coded from that measurement rather than recomputed per facility, unnormalized to rated load, and untested against the smoothing that many asynchronous servers may provide at facility scale. The gas complete-site figure excludes redundancy, forced outages, gas interconnection, emissions-control permitting, switchgear, and land, all of which raise it. The 45V implementation covers a single 2030 cohort rather than a year-by-year project pipeline. The investment inversion treats learning rates and installed bases for the conversion technologies as flagged estimates. None of these limitations, in the assessment of the external review the model underwent, overturns the central conclusions; several of them—the gigawatt-scale facility ladder above all—define the next revision.

## 7. Conclusion

Four directions define the next revision, in priority order. First, the facility ladder must extend to the gigawatt class, where the marginal project now lives: at that scale on-site gas becomes a dedicated plant with its own interconnection and emissions permitting, the renewable-natural-gas ceiling binds proportionally harder, and the hydrogen fuel chain shifts from trailered delivery to dedicated pipeline, moving its floor. Second, the gas path should be replaced by a fundamentals model trained on storage, production, export capacity, and weather—the machinery exists in the author's forecasting work and its absence here is the model's weakest link. Third, the ramp statistic should be recomputed live from the traces per facility configuration, normalized to rated load, and tested against the smoothing that aggregating many asynchronous training jobs may provide, which could shrink the buffer further or, for synchronized large runs, enlarge it. Fourth, the 45V implementation should move from a single cohort to a year-by-year project pipeline, which matters for any technology trajectory that leans on the credit's timing.

Measured against a grid whose price is inflated by the very demand growth that motivates the question, no on-site supply technology powers a data center more cheaply than the wire, once the retained grid relationship is priced. Gas comes closest and stays twenty-odd dollars short; storage is a capped service with marginal-gas carbon; hydrogen from grid electricity fails twice over; and the low-carbon fuels that work today are resource-capped near a sixth of site energy. Investment changes exactly one of these facts: roughly a trillion dollars of global electrolyser deployment, sited on curtailed power below twenty dollars per megawatt-hour, buys PEM hydrogen power at 1.9 times the grid (2.3 for the engine) with a carbon reduction of about 85 percent and no resource ceiling—depth, not parity, because parity is not for sale at any budget. The rational program that follows is short: right-size and run continuously, which is free; open capacity markets to behind-the-meter assets, which is free; price interconnection access explicitly, because it is the product actually being bought; fund electrolyser deployment on curtailment if depth is required; and spend nothing on conversion-hardware cost reduction, whose free-hardware floor already exceeds the grid. On-site generation for data centers is an access and depth product. Priced as one, most of the current investment conversation is aimed at the wrong term.

*Note on the reference list. Entries [4], [15], and [26] are cited in the generic form carried by the model's parameter files; their exact report titles were not re-verified against the publishers during preparation and should be confirmed before submission. All other entries were verified against the cited documents.*